\documentclass[10pt]{article}
\usepackage{cite}
\author{Fr\'ed\'eric Bucci$^{1,2}$, Iacopo Mastromatteo$^{2,3}$\\
Michael Benzaquen$^{2,3,4}$ and Jean-Philippe Bouchaud$^{2,3,5}$\\}
\date{\small%
$^1$ \textit{Scuola Normale Superiore, Piazza dei Cavalieri 7, 56126 Pisa, Italy} \\%
$^2$ \textit{Chair of Econophysics and Complex Systems, Ecole polytechnique, 91128 Palaiseau Cedex, France} \\%
$^3$ \textit{Capital Fund Management, 23-25, Rue de l'Universit\'e 75007 Paris, France}  \\%
$^4$ \textit{Ladhyx  UMR CNRS 7646 \& Department of Economics, Ecole polytechnique, \\ 91128 Palaiseau Cedex, France} \\%
$^5$ \textit{CFM-Imperial Institute of Quantitative Finance, Department of Mathematics, Imperial College, 180 Queen's Gate, London SW7 2RH}\\%
\today}
\title{Impact is not just volatility}
\usepackage{booktabs}
\usepackage{amsmath,amssymb}
\usepackage{amssymb}
\usepackage{graphicx}
\usepackage{amsmath}
\usepackage{amsfonts}
\usepackage{geometry}
\usepackage{bm}
\usepackage{breqn}
\usepackage{bbm}
\usepackage{textcomp}
\usepackage{xcolor}
\usepackage{subfig}
\usepackage{amsmath}
\usepackage{amssymb}
\usepackage{geometry}
\usepackage{mathtools}
\usepackage{bbm}
\usepackage{tabularx}

\newcommand{\imp}{\mathcal{I}}

\begin{document}
% Hint: \title{what ever}, \author{who care} and \date{when ever} could stand 
% before or after the \begin{document} command 
% BUT the \maketitle command MUST come AFTER the \begin{document} command! 

\maketitle

\begin{abstract}
The notion of market impact is subtle and sometimes misinterpreted. Here we argue that impact should not be misconstrued as volatility. 
In particular, the so-called ``square-root impact law'', which states that impact grows as the square-root of traded volume, has nothing to do
with price diffusion, i.e. that typical price changes grow as the square-root of time. We rationalise empirical findings on impact and volatility 
by introducing a simple scaling argument and confronting it to data.
\end{abstract}\medskip

Transaction Cost Analysis has become a very important issue in the Asset Management industry. Transaction costs account
for a substantial part of the profits or losses of any investment strategy. When executing an order to buy or to sell, 
investors and trading firms have to worry about several sources of costs. Some costs are easy to identify and quantify, like market fees or 
spread costs. Much more subtle, but dominant for large portfolios, are {\it impact costs}. Intuitively, market impact describes the
fact that, {\it on average}, buy orders tend to push the price up and sell orders tend to drag the price down. All the subtlety, however, 
lies in the words ``on average''. Clearly, while our putative investor is executing his/her buy order, many things can happen: other investors 
may simultaneously buy or sell, market makers/high frequency traders may unload their inventories, or some news may become available, pushing the 
price up or down. While some of these events may be directly related to his/her buy order, most of them result in a price move unbeknownst to our investor. 

For a large enough number of executed orders, these random price moves average to zero. But for any given order executed within a time $T$, 
the price will randomly move up or down by an amount $\sim \sigma \sqrt{T}$, where $\sigma$ is the volatility. The impact of an order, 
on the other hand, is the part of the price move that survives upon averaging. Not surprisingly, this impact is much smaller than $\sigma \sqrt{T}$ 
for small order sizes -- see below for more precise statements. This definition of impact should be further refined to distinguish between the 
reactional (or mechanical) impact, that would exist even for trades without any 
information content, and the prediction related impact that reveals the information content of the trade, see \cite{TQP}. The latter component is usually very small for
medium to long term investors, since information (if any) is supposed to affect investment time scales much longer than the execution time $T$ 
(typically several weeks or months compared to $T \sim$ a few days at most).\footnote{Most trades in the ANcerno database discussed below appear to belong to medium or long term investors.} 

One interesting question concerns the dependence of the reactional impact $\imp(Q, T)$ on the size $Q$ and duration $T$ of the executed order. A now 
commonly accepted result is the so-called square-root law (see e.g. \cite{Torre,Almgren,Engle,Moro,Toth,Brokmann,Zarinelli,Bacry,Bonart,Toth2,AQR,Bucci}) which states that in normal trading conditions
\[
\imp(Q, T) \approx Y \sigma \sqrt{\frac{Q}{V}},
\]
where $Y$ is a constant of order unity, and $\sigma$ and $V$ are, respectively, the daily volatility and daily volume corresponding to the traded asset.
In fact, a more accurate description was proposed theoretically in \cite{Donier, Benzaquen}, where this square-root dependence becomes linear for small $Q$; more precisely:
\[
\imp(Q, T) = \sigma \sqrt{\frac{Q}{V}}\, {\cal F}(\phi), \qquad \phi:=\frac{Q}{VT},
\]
where $T$ is expressed in days such that $VT$ is the total volume executed during $T$. 
The scaling function ${\cal F}(\phi)$ is monotonic and behaves as $\sqrt{\phi}$ for $\phi \to 0$ and as a constant $Y$ for $\phi \to \infty$. 
Therefore, $\imp(Q, T)$ is linear in $Q$ for small $Q$ at fixed $T$, and crosses over to a square-root for large $Q$.\footnote{Note that, as discussed in \cite{TQP}, this behaviour
is not expected to hold in extreme trading conditions, for example when $\phi$ is large and $T$ is small. If latent liquidity has no time to reveal itself in the order book, convex impact or event runaway situations can ensue -- see for example \cite{dellAmico}.}
This prediction appears to describe empirical data surprisingly well, see \cite{BBLB}.

Such a functional form for the reaction impact has two immediate consequences. One is that for $Q \ll VT$ (i.e. when the volume of the whole order $Q$ is 
much smaller than the market volume during time $T$), reaction impact is much smaller than the typical price moves: $\imp(Q, T) \ll \sigma \sqrt{T}$. In other 
words, the average impact (and therefore the associated impact cost) incurred by our investor is very small compared to the uncertainty on the price move 
during execution. Large trading firms with active investment strategies will however be mostly sensitive to the former (as the latter averages to zero), whereas one-off investors
will want to minimise uncertainty, for fear of an adverse price move while their order is executed. The latter concern is at the heart of the famous Almgren-Chriss formalism \cite{Almgren-Chriss}.

The second, somewhat surprising consequence is that in the square-root regime, reaction impact depends only weakly on the execution time $T$ -- whereas of 
course the typical price changes increase as $\sqrt{T}$. Now, this $\sqrt{T}$ dependence was recently argued to be the mechanism at the heart 
of the square-root impact law \cite{Capponi}. The argument, in a nutshell, is that typical investors are essentially sensitive to price uncertainty, so perceived costs 
behave as $\sigma \sqrt{T}$. But if their order of size $Q$ is executed at a constant rate $\varphi$, the time needed to complete the order is $T = Q/\varphi$. Hence
apparent impact behaves as $\sigma \sqrt{Q/\varphi}$, i.e. a square-root dependence on $Q$. We believe that this argument is very misleading, and  
fails at explaining why reaction impact (i.e. the {\it average} of price moves, and not its root-mean-square) behaves as $\sqrt{Q}$.\footnote{More generally, the average overhead in costs due to impact $\mathcal{C}$ can be directly related to $\imp(Q, T) $ via the relation $\mathcal{C} = \int_0^T \, {\mathrm d}t \, \dot Q(t) \imp(Q(t), t)$, where $Q(t)$ is the quantity executed by the investor at each time $t \leq T$. On the other hand, the idiosyncratic price moves $\sim \sigma \sqrt{Q/\varphi}$ only relate to execution risk, and are not directly linked to the average cost $\mathcal{C} $ paid by investors, for they average out after a sufficiently large number of investment decisions are executed.}

\begin{figure}
\begin{minipage}{1.0\textwidth}
\centering
\includegraphics[width=1.0\linewidth]{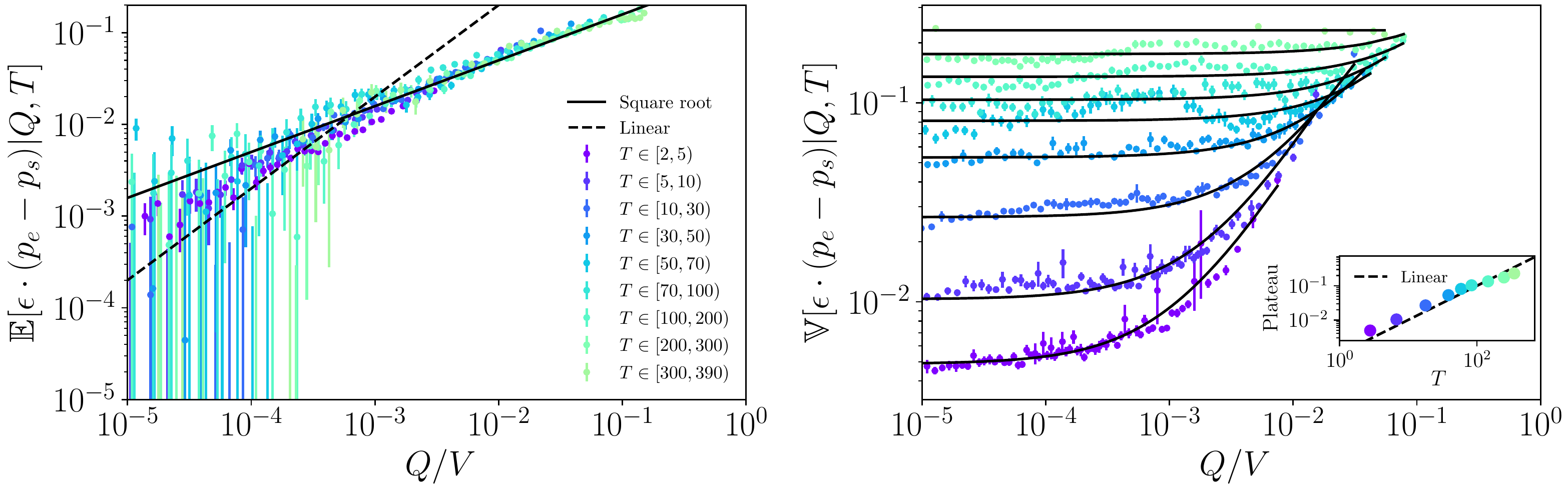}
\end{minipage}
\caption{Market impact curves $\imp(Q, T) = \mathbb{E}[\epsilon \cdot (p_{\textrm{e}}-p_{\textrm{s}})|Q, T]$ (left panel) and price uncertainty measured by $\mathbb{V}[\epsilon \cdot (p_{\textrm{e}}-p_{\textrm{s}})|Q, T]$ (right panel) as a function of the relative metaorder size $Q/V$ for different buckets of order duration $T$: with an abuse of notation the log-prices $p_{\textrm{e},\textrm{s}}$ are rescaled by the average daily volatility per stock while the order size is given by the ratio between the number of shares $Q$ and the average daily total market volume $V$ per stock.  As shown in the inset, the plateau value increases linearly with the duration $T$, as expected for a random walk.}
\label{fig1}
\end{figure}

The difference between these two quantities is shown in Fig. 1, based on the ANcerno database (for full details, see refs \cite{Bucci, BBLB, SlowBucci}). In the left panel, we show the average price difference 
between the start and the end of the order execution, conditionned to the order size $Q$, and for different order durations $T$ (different colors). 
One clearly sees the crossover between a $T$-dependent, steeper than square-root regime for small $Q/V$ and a $T$-independent, square-root regime for larger $Q/V$, as already reported in \cite{Zarinelli, Bucci, BBLB, SlowBucci}. Note that over $80\%$ of the empirical data lies in the square-root regime. 

In the right panel, we show the variance of price differences, again conditionned to the order fraction and for different order durations. For small
$Q$, this variance is nearly independent of the order size $Q$ but linearly increases with $T$ as expected (see inset). For larger $Q$, the variance acquires some dependence 
on $Q/V$, specially for small $T$. In order to rationalize these findings, let us postulate that the price change $\Delta p$ between the start and the end of the 
execution of an order of duration $T$ is given by the sum of an impact contribution and a volatility contribution, i.e. 
\[
\Delta p=p_{\textrm{e}}-p_{\textrm{s}}= \epsilon \cdot \imp(Q, T) \times (1+a\,\eta)  +\sigma \sqrt{T} \xi,
\]
where $\epsilon=\pm 1$ depending on the sign of the order (buy or sell), $a$ is a certain fitting parameter, $\eta, \, \xi$ are two independent random variables with zero mean and unit variance and $T$ is, as above, measured as a fraction of the trading day. From this ansatz, it follows that:
\[
\mathbb{E}[\epsilon \cdot \Delta p|Q, T] = \imp(Q, T) ,
\]
as it should be of course, and 
\[
  \mathbb{V}[\epsilon \cdot \Delta p|Q, T] = \sigma^2 T \left( 1 + a^2 \phi \, {\cal F}^2(\phi)  \right).
\]
This prediction is plotted in the right panel of Fig. 1, with $a$ as the only fitting parameter ($a \approx 10^{-1}$).
%This prediction is plotted in the right panel of Fig. 1, with $a$ as the only fitting parameter (corresponding to $a \approx 10^{-2}$***really ? I would prefer $a=0.1$, isn't there a square somewhere???**). 
%\clearpage
Taken together, our ansatz and the two panels of Fig. 1 confirm that:
\begin{enumerate}
\item The square-root impact law for $\mathbb{E}[\epsilon \cdot \Delta p|Q, T]$ is completely unrelated to the scaling of the volatility as $\sqrt{T}$;
\item The square-root impact law and its fluctuations (parameterized by $a$) allows one to understand the systematic increase of volatility in the presence of a locally large order (large $\phi$);
\item As expected, price uncertainty (measured by $\mathbb{V}[\epsilon \cdot \Delta p|Q, T]$) largely exceeds the average reaction impact contribution: compare the square of the y-axis of the left panel with the y-axis of the right panel. This means, as emphasized in \cite{Capponi}, that impact has a poor explanatory power compared to volatility. If an asset manager represents 
$1\%$ of the market volume, he or she contributes to $5\%$ of the volatility, which gives an $R^2$ for the impact term of $\approx 2.5 \times 10^{-3}$.
\end{enumerate}

In conclusion, we wanted in this short note to point out some basic facts about impact that are sometimes misinterpreted, with Ref.~\cite{Capponi} as a case in point. In particular, even when execution risk is relevant for some investors (and at the core of the Almgren-Chriss formalism \cite{Almgren-Chriss}), price variance should certainly not be misconstrued as price impact. 

\section*{Acknowledgments}

This research was conducted within the \emph{Econophysics \&  Complex Systems} Research Chair, under the aegis of the Fondation du Risque, the Fondation de l'Ecole polytechnique, the Ecole polytechnique and Capital Fund Management. We thank F. Lillo and J. Gatheral for sharing their thoughts with us.  

\section*{Data availability statement}
The data were purchased by Imperial College from the company ANcerno Ltd (formerly the Abel Noser Corporation) which is a widely recognised consulting firm that works with institutional investors to monitor their equity trading costs. Its clients include many pension funds and asset managers. The authors do not have permission to redistribute them, even in aggregate form. Requests for this commercial dataset can be addressed directly to the data vendor.  See www.ancerno.com for details.


\begin{thebibliography}{100}
\bibitem{TQP}J.-P. Bouchaud, J. Bonart, J. Donier, \& M. Gould, \textit{Trades, Quotes and Prices: Financial Markets Under The Microscope} Cambridge University Press, (2018). 
\bibitem{Torre}N. Torre, \textit{Barra market impact model handbook.} BARRA Inc., Berkeley, 1997.
\bibitem{Almgren}R. Almgren, C. Thum, E. Hauptmann, \& H. Li, \textit{Direct estimation of equity market impact.} Risk, vol. 57, (2005).
\bibitem{Engle}R. F. Engle, R. Ferstenberg, \& J. Russell, \textit{Measuring and modeling execution cost and risk.} Chicago GSB Research Paper, no. 08-09, (2008).
\bibitem{Moro} {E. Moro, J. Vicente, L.G. Moyano, A. Gerig, J.D. Farmer,
G. Vaglica, F. Lillo, \& R.N. Mantegna}, \textit{Market impact and trading profile of hidden orders in stock markets}. Phys. Rev. E  80, 066102 (2009).
\bibitem{Toth}B. T\'oth, Y. Lemperiere, C. Deremble, J. De Lataillade, J. Kockelkoren, \&
J.-P. Bouchaud, \textit{Anomalous price impact and the critical nature of liquidity in financial markets.} Physical Review X, vol. 1, no. 2, p. 021006, (2011).
\bibitem{Brokmann} Brokmann, X., Serie, E., Kockelkoren, J., \& Bouchaud, J. P. \textit{Slow decay of impact in equity markets}. Market Microstructure and Liquidity, 1(02), 1550007 (2015).
\bibitem{Zarinelli} E. Zarinelli, M. Treccani, J. D. Farmer \& F. Lillo, \textit{Beyond the Square Root: Evidence for Logarithmic Dependence of Market Impact on Size and Participation Rate}. Market Microstructure and Liquidity, Vol. 1, No. 2, (2015).
\bibitem{Bacry}E. Bacry, A. Iuga, M. Lasnier, \& C. A. Lehalle, C. A., \textit{Market impacts and the life cycle of investors orders}. Market Microstructure and Liquidity, 1(02), 1550009, (2015).
\bibitem{Bonart} J. Donier, \& J. Bonart, \textit{ A Million Metaorder Analysis of Market Impact on the Bitcoin}, Market Microstructure and Liquidity, Vol. 01, No. 02, (2015).
\bibitem{Toth2}B. Toth, Z. Eisler, \& J.-P. Bouchaud, \textit{The square-root impact law also holds for option markets}, Wilmott (2016).
\bibitem{AQR} Frazzini, A., Israel, R., \& Moskowitz, T. J. (2018). Trading costs. Available at SSRN: https://ssrn.com/abstract=3229719. 
\bibitem{Bucci} F. Bucci, I. Mastromatteo, Z. Eisler, F. Lillo, J.-P. Bouchaud, \& C.-A. Lehalle, \textit{Co-impact: Crowding effects in institutional trading activity}, https://arxiv.org/abs/1804.09565, (2018).
\bibitem{Donier} J. Donier, J. Bonart, I. Mastromatteo, \& J.-P. Bouchaud, \textit{A fully consistent, minimal model for non-linear market impact}, Quantitative Finance 15, 1009-1121, (2015).
\bibitem{Benzaquen} M. Benzaquen, \& J.-P. Bouchaud, \textit{Market impact with multi-timescale liquidity}, Quantitative Finance, 2018, p. 1-10, (2018).
\bibitem{dellAmico} Dall'Amico, L., Fosset, A., Bouchaud, J. P., \& Benzaquen, M. (2019). How does latent liquidity get revealed in the limit order book? Journal of Statistical Mechanics: Theory and Experiment, 2019(1), 013404.
\bibitem{BBLB} F. Bucci, M. Benzaquen, F. Lillo,  \& J.-P. Bouchaud, \textit{Crossover from linear to square-Root market impact}, Physical Review Letters, 122(10), 108302, (2019).
\bibitem{Almgren-Chriss} Almgren, R., \& Chriss, N. (2001). Optimal execution of portfolio transactions. Journal of Risk, 3, 5-40.
\bibitem{Capponi} F. Capponi \& R. Cont, \textit{Trade Duration, Volatility and Market Impact}, Available at SSRN https://ssrn.com/abstract=3351736, (2019).
\bibitem{SlowBucci}F. Bucci, M. Benzaquen, F. Lillo \& J.-P. Bouchaud, \textit{Slow decay of impact in equity markets: insights from the ANcerno database}, https://arxiv.org/abs/1901.05332, (2019).
\end{thebibliography}
\end{document}